\documentclass{llncs}
\usepackage[ruled, vlined, nofillcomment, linesnumbered]{algorithm2e}
\usepackage{cite}
\usepackage{url}
\usepackage{amsmath}
\usepackage{amssymb}
\usepackage{graphicx}
\usepackage{subfig}
\usepackage[english]{babel}
\usepackage{booktabs}
\usepackage{nicefrac}
\usepackage{verbatim}
\usepackage{color}
\usepackage{float}
\usepackage{xspace}
\usepackage{mathtools}
\usepackage{etex}

\graphicspath{ {images/} }

\SetKwFunction{update}{Update}

\newcommand{\spara}[1]{{\smallskip\noindent{\bf {#1}}}}

\newcommand{\nevents}{\ensuremath{k}\xspace}
\newcommand{\metagraph}{interaction meta-graph\xspace}

\newcommand{\set}[1]{\left\{#1\right\}}

\newcommand{\abs}[1]{{\left|#1\right|}}

\newcommand{\real}{\mathbb{R}}
\newcommand{\np}{\textbf{NP}}
\newcommand{\naturals}{\mathbb{N}}

\newcommand{\bigO}{\ensuremath{\mathcal{O}}\xspace}

\newcommand{\graph}{\ensuremath{H}\xspace}
\newcommand{\node}{\ensuremath{u}\xspace}
\newcommand{\nnodes}{\ensuremath{n}\xspace}
\newcommand{\nodes}{\ensuremath{N}\xspace}
\newcommand{\edges}{\ensuremath{{\mathcal{I}}}\xspace}

\newcommand{\info}{\ensuremath{\alpha}\xspace}
\newcommand{\nedges}{\ensuremath{m}\xspace}
\newcommand{\ntopics}{\ensuremath{L}\xspace}

\newcommand{\mgraph}{\ensuremath{G}\xspace}
\newcommand{\mnodes}{\ensuremath{\edges}\xspace}
\newcommand{\medges}{\ensuremath{E}\xspace}
\newcommand{\mweight}{\ensuremath{c}\xspace}
\newcommand{\nodeweight}{\ensuremath{p}\xspace}
\newcommand{\none}{\ensuremath{i}\xspace}
\newcommand{\ntwo}{\ensuremath{j}\xspace}
\newcommand{\event}{\ensuremath{T}\xspace}
\newcommand{\events}{\ensuremath{\mathcal{T}}\xspace}

\newcommand{\treeroot}{\ensuremath{r}\xspace}
\newcommand{\tbudget}{\ensuremath{I}\xspace}
\newcommand{\distbudget}{\ensuremath{B}\xspace}
\newcommand{\enodes}{\ensuremath{V_e}\xspace}
\newcommand{\eedges}{\ensuremath{E_e}\xspace}
\newcommand{\twindow}{\ensuremath{[s,f]}\xspace}

\newcommand{\eupperbound}{\ensuremath{U}\xspace}
\newcommand{\forest}{\ensuremath{F}\xspace}
\newcommand{\rooteddag}{\ensuremath{D}\xspace}
\newcommand{\inedges}{\ensuremath{\delta^{+}}\xspace}

\newcommand{\charikaralgo}{\ensuremath{CharikarDST}\xspace}
\newcommand{\charikarlevel}{\ensuremath{\ell}\xspace}
\newcommand{\charikarterminals}{\ensuremath{X}\xspace}
\newcommand{\quota}{\ensuremath{Q}\xspace}
\newcommand{\lquota}{\ensuremath{Q_l}\xspace}
\newcommand{\uquota}{\ensuremath{Q_u}\xspace}

\newcommand{\greedy}{{\it greedy\/}\xspace}
\newcommand{\lst}{{\it DP\/}\xspace}
\newcommand{\lstdij}{{\it DP+dij\/}\xspace}
\newcommand{\aquota}{{\it binary\_search\/}\xspace}
\newcommand{\random}{{\it random\/}\xspace}

\newcommand{\upperbound}{{\it upperbound\/}\xspace}

\newcommand{\pcsd}{\ensuremath{PCSD}\xspace}

\DeclarePairedDelimiter\floor{\lfloor}{\rfloor}
\DeclareMathOperator*{\argmin}{arg\,min}
\DeclareMathOperator*{\argmax}{arg\,max}

\newcommand{\tpcst}{\ensuremath{\text{\sc TMax\-Tree}}\xspace}
\newcommand{\pcst}{\ensuremath{\text{\sc Max\-Tree}}\xspace}
\newcommand{\ktrees}{\ensuremath{k\text{-{\sc Max\-Trees}}}\xspace}

\newcommand{\enron}{\textit{Enron}\xspace}
\newcommand{\beefban}{\textit{\#beefban}\xspace}
\newcommand{\ukraine}{\textit{\#ukraine}\xspace}
\newcommand{\baltimore}{\textit{\#baltimore}\xspace}

\usepackage{tikz,pgfplots,pgfplotstable}
\usetikzlibrary{positioning,fit,shapes}
\usetikzlibrary{decorations.pathreplacing}
\usetikzlibrary{arrows}

\pgfdeclarelayer{background}
\pgfdeclarelayer{foreground}
\pgfsetlayers{background,main,foreground}

\makeatletter
\tikzset{multicircle/.style  args={#1, #2}{ alias=tmp@name,   postaction={    insert path={
     \pgfextra{     \pgfpointdiff{\pgfpointanchor{\pgf@node@name}{center}}                  {\pgfpointanchor{\pgf@node@name}{east}}     \pgfmathsetmacro\insiderad{\pgf@x}             \fill[white] (\pgf@node@name.center)  circle (\insiderad-\pgflinewidth);        \draw[#2] (\pgf@node@name.center)  circle (\insiderad-\pgflinewidth);        \fill[#2] (\pgf@node@name.center)  -- ++(0:\insiderad-\pgflinewidth) arc (0:#1:\insiderad-\pgflinewidth)--cycle;        }}}}}
\makeatother

\definecolor{yafaxiscolor}{rgb}{0.3, 0.3, 0.3}
\definecolor{yafcolor1}{rgb}{0.4, 0.165, 0.553}
\definecolor{yafcolor2}{rgb}{0.949, 0.482, 0.216}
\definecolor{yafcolor3}{rgb}{0.47, 0.549, 0.306}
\definecolor{yafcolor4}{rgb}{0.925, 0.165, 0.224}
\definecolor{yafcolor5}{rgb}{0.141, 0.345, 0.643}
\definecolor{yafcolor6}{rgb}{0.965, 0.933, 0.267}
\definecolor{yafcolor7}{rgb}{0.627, 0.118, 0.165}
\definecolor{yafcolor8}{rgb}{0.878, 0.475, 0.686}
\definecolor{yafcolor9}{rgb}{0.965, 0.733, 0.767}

\newlength{\yafaxispad}
\newlength{\yaftlpad}
\newlength{\yaflabelpad}
\newlength{\yafaxiswidth}
\newlength{\yafticklen}
\makeatletter
\def\pgfplots@drawtickgridlines@INSTALLCLIP@onorientedsurf#1{}
\makeatother

\newcommand{\yafdrawxaxis}[2]{
  \pgfplotstransformcoordinatex{#1}\let\xmincoord=\pgfmathresult 
  \pgfplotstransformcoordinatex{#2}\let\xmaxcoord=\pgfmathresult 
  \pgfsetlinewidth{\yafaxiswidth} 
  \pgfsetcolor{yafaxiscolor}
  \pgfpathmoveto{\pgfpointadd{\pgfpointadd{\pgfplotspointrelaxisxy{0}{0}}{\pgfqpointxy{\xmincoord}{0}}}{\pgfqpoint{-0.5\yafaxiswidth}{\yafaxispad}}}
  \pgfpathlineto{\pgfpointadd{\pgfpointadd{\pgfplotspointrelaxisxy{0}{0}}{\pgfqpointxy{\xmaxcoord}{0}}}{\pgfqpoint{0.5\yafaxiswidth}{\yafaxispad}}}
  \pgfusepath{stroke}

}
\newcommand{\yafdrawyaxis}[2]{
  \pgfplotstransformcoordinatey{#1}\let\ymincoord=\pgfmathresult 
  \pgfplotstransformcoordinatey{#2}\let\ymaxcoord=\pgfmathresult 
  \pgfsetlinewidth{\yafaxiswidth} 
  \pgfsetcolor{yafaxiscolor}
  \pgfpathmoveto{\pgfpointadd{\pgfpointadd{\pgfplotspointrelaxisxy{0}{0}}{\pgfqpointxy{0}{\ymincoord}}}{\pgfqpoint{\yafaxispad}{-0.5\yafaxiswidth}}}
  \pgfpathlineto{\pgfpointadd{\pgfpointadd{\pgfplotspointrelaxisxy{0}{0}}{\pgfqpointxy{0}{\ymaxcoord}}}{\pgfqpoint{\yafaxispad}{0.5\yafaxiswidth}}}
  \pgfusepath{stroke}
}

\pgfplotscreateplotcyclelist{yaf}{{yafcolor1,mark options={scale=0.75},mark=o}, 
{yafcolor2,mark options={scale=0.75},mark=square},
{yafcolor3,mark options={scale=0.75},mark=triangle},
{yafcolor4,mark options={scale=0.75},mark=o},
{yafcolor5,mark options={scale=0.75},mark=o},
{yafcolor6,mark options={scale=0.75},mark=o},
{yafcolor7,mark options={scale=0.75},mark=o},
{yafcolor8,mark options={scale=0.75},mark=o}} 

\pgfplotsset{axis y line=left, axis x line=bottom,
  tick align=outside,
  compat = 1.3,
  tickwidth=\yafticklen,
  clip = false,
  every axis title shift = 0pt,
    x axis line style= {-, line width = 0pt, opacity = 0},
    y axis line style= {-, line width = 0pt, opacity = 0},
    x tick style= {line width = \yafaxiswidth, color=yafaxiscolor, yshift = \yafaxispad},
    y tick style= {line width = \yafaxiswidth, color=yafaxiscolor, xshift = \yafaxispad},
    x tick label style = {font=\scriptsize, yshift = \yaftlpad},
    y tick label style = {font=\scriptsize, xshift = \yaftlpad},
    every axis y label/.style = {at = {(ticklabel cs:0.5)}, rotate=90, anchor=center, font=\scriptsize, yshift = -\yaflabelpad},
    every axis x label/.style = {at = {(ticklabel cs:0.5)}, anchor=center, font=\scriptsize, yshift = \yaflabelpad},
    x tick label style = {font=\scriptsize, yshift = 1pt},
    grid = major,
    major grid style  = {dash pattern = on 1pt off 3 pt},
  every axis plot post/.append style= {line width=\yafaxiswidth} ,
  legend cell align = left,
  legend style = {inner sep = 1pt, cells = {font=\scriptsize}},
  legend image code/.code={    \draw[mark repeat=2,mark phase=2,#1] 
    plot coordinates { (0cm,0cm) (0.15cm,0cm) (0.3cm,0cm) };  } 
}

\begin{document}

\title{Discovering topically- and temporally-coherent events in interaction networks}

\author{}

\author{Han Xiao$^1$ \and Polina Rozenshtein$^2$ \and Aristides Gionis$^2$}

\institute{
$^1$Helsinki Institute for Information Technology and \\
Department of Computer Science \\
University of Helsinki, Finland \\
\url{hxiao@cs.helsinki.fi} \\
\vspace{1em}
$^2$Helsinki Institute for Information Technology and \\
Department of Computer Science \\
Aalto University, Finland \\   
\url{firstname.lastname@aalto.fi}}

\maketitle

\begin{abstract}

With the increasing use of online communication platforms, 
such as email, Twitter, and messaging applications, 
we are faced with a growing amount of data that combine
{\em content} (what is said), 
{\em time} (when), and {\em user} (by whom) information. 
Discovering meaningful patterns and 
understand what is happening in this data is an important challenge. 
We consider the problem of mining online communication data 
and finding top-$\nevents$ {\em temporal events}. 
A temporal event is a coherent topic that is discussed frequently
in a relatively short time span, 
while its information flow respects the underlying network.

Our method consists of two steps. 
We first introduce the notion of \textit{\metagraph}, 
which connects associated interactions. 
Using this notion, we define a \textit{temporal event} to be a subset of interactions that 
($i$) are topically and temporally close and 
($ii$) correspond to a tree that captures the information flow. 
Finding the best temporal event 
leads to a budget version of the prize-collecting Steiner-tree (PCST) problem, 
which we solve using three different methods: 
a greedy approach, 
a dynamic-programming algorithm, 
and an adaptation to an existing approximation algorithm. 
Finding the top-$\nevents$ events 
maps to a maximum set-cover problem, and thus, solved by greedy algorithm. 
We compare and analyze our algorithms in both synthetic and real datasets, 
such as Twitter and email communication. 
The results show that our methods are able to detect meaningful temporal events. 
\end{abstract}

\section{Introduction}
\label{sec:intro}

Event detection is a fundamental data-mining problem 
in many different domains, such as, 
time series and data streams \cite{guralnik1999event}, 
point clouds and vector spaces \cite{breunig2000lof}, and 
networks \cite{boden2012mining}.  
In this paper we focus on the problem of detecting events in networks,  
in particular, networks that contain both \emph{content} and \emph{time} information.
An \textit{interaction} 
$(u, v, \info, t)$ occurs whenever a piece of information $\info$ is
exchanged between two network entities $u$ and $v$ at time $t$.
Examples of interaction networks include data communication networks, 
such as email, Twitter, or online messaging systems.

Our goal is to summarize the network activity 
by finding the top-$\nevents$ events.
We consider an interaction network $\graph=(\nodes, \edges)$, 
where interactions {\edges} 
take place among a set of network entities {\nodes}. 
The interactions in {\edges} are directed, annotated with content information, 
and time-stamped. 
We define an \textit{event} in the interaction graph $\graph$ 
to be a subset of interactions, $\edges' \subseteq \edges$ that are 
($i$) temporally close, 
($ii$) topically similar, and 
($iii$) correspond to a tree that captures the information flow in the network. 
The intuition behind representing events as trees is similar to the work by Yang \cite{yang2010modeling}. 

We convert the interaction network $\graph=(\nodes, \edges)$ 
into a weighted \emph{\metagraph} $\mgraph=(\mnodes, \medges)$,
that is, a graph whose vertices are the interactions~\edges. Two interactions $\none, \ntwo \in \mnodes$ are connected in  \mgraph
if it is possible to explain the information flow between \none and \ntwo.
In particular, we consider three types of flow: 
\textit{broadcast}, 
\textit{relay} 
and \textit{reply}. 
The edge weights of the \metagraph \mgraph 
measure the topic dissimilarity between connected interactions. 
Our transformation from the interaction network 
to the  \metagraph has the interesting property
that an \textit{event} in the interaction graph $\graph$ 
corresponds to a tree $\event$ in the \metagraph \mgraph. 
The root of the tree $\event$ is interpreted as the source of the event. 
Downstream interactions (interactions that are reachable from the root)
are due to information propagation.

\begin{figure}[t]
\begin{tikzpicture}[scale=0.8,every node/.style={scale=0.8}]]

\tikzstyle{personbox} = [font=\small, draw = black!50, rectangle, inner sep = 8pt, text=black!70, rounded corners, text centered]
\tikzstyle{communication} = [thick]
\tikzstyle{interaction} = [font=\small, thick, circle, inner sep = 3pt, text= black!70]
\tikzstyle{iedge} = [thick]

\tikzset{
  text style/.style={text=black!70, font=\footnotesize}
}

\node[personbox] (ceo) at (5,5) {CEO};
\node[personbox] (pm) at (2.5,3) {Project manager (PM)};
\node[personbox] (tm1) at (-1,1) { Team member 1 (TM1)};
\node[personbox] (tm2) at (5,-0.5) {Team member 2 (TM2)};

\draw (ceo.west) 
	edge [yafcolor4!80, communication, ->, bend right = 20] 
	node [text style, left, xshift=-5pt] {(1,`progress',Mon)} 
	([xshift=-10pt]pm.north);
\draw (pm) 
	edge [yafcolor4!80, communication, ->] 
	node [text style, xshift=8pt] {(4,`progress',Thu)} 
	(ceo);
\draw ([xshift=50pt]pm.north) 
	edge [draw=yafcolor3!80, communication, ->, bend right = 20] 
	node [text style, right, yshift=-10pt, xshift=-6pt] {(6,`suggestion',Thu)} 
	(ceo.south);
\draw (pm) 
	edge [yafcolor4!80, communication, ->, bend right = 10] 
	node [text style, left, xshift=-5pt] {(2,`progress',Tue)} 
	(tm1);
\draw (pm) 
	edge [yafcolor4!80, communication, ->, bend left = 10] 
	node [text style, right, xshift=2pt] {(2,`progress',Tue)} 
	(tm2);
\draw (tm2) 
	edge [yafcolor4!80, communication, ->, bend left = 10] 
	node [text style, left, yshift=-22pt, xshift=17pt] {(3,`progress',Wed)} 
	(pm);
\draw (tm1) 
	edge [draw=yafcolor3!80, communication, ->, bend right = 10] 
	node [text style, yshift=-5pt, xshift=10pt] {(5,`suggestion',Wed)} 
	(pm);
\draw (tm1) 
	edge [draw=yafcolor5!80, communication, ->, bend right = 10] 
	node [text style, below, xshift=-30pt] {(7,`football',Fri)} 
	(tm2);

\node[text=black!70, font=\large] at (2.5,-2) {(a)};

\node[interaction, draw=yafcolor4!80, fill=yafcolor4!05] (i1) at (11,5) {1};
\node[interaction, draw=yafcolor4!80, fill=yafcolor4!05] (i2) at (9.9,4.1) {2};
\node[interaction, draw=yafcolor4!80, fill=yafcolor4!05] (i3) at (9,3) {3};
\node[interaction, draw=yafcolor4!80, fill=yafcolor4!05] (i4) at (12.5,2) {4};
\node[interaction, draw=yafcolor3!80, fill=yafcolor3!05] (i5) at (11.6,0.9) {5};
\node[interaction, draw=yafcolor3!80, fill=yafcolor3!05] (i6) at (10.5,0) {6};
\node[interaction, draw=yafcolor5!80, fill=yafcolor5!05] (i7) at (9.2,0.8) {7};

\draw (i1) edge [thick, ->, bend right = 10, draw=yafcolor1!60] (i2);
\draw (i1) edge [thick, ->, bend left = 10,  draw=yafcolor2!60] (i4);
\draw (i1) edge [thick, dashed, ->, bend left = 10,  draw=yafcolor2!60] (i6);
\draw (i2) edge [thick, ->, bend right = 10, draw=yafcolor2!60] (i3);
\draw (i2) edge [thick, ->, bend left = 10,  draw=yafcolor7!60] (i4);
\draw (i2) edge [thick, dashed, ->, bend left = 0,   draw=yafcolor2!60] (i5);
\draw (i2) edge [thick, dashed, ->, bend right = 10, draw=yafcolor7!60] (i6);
\draw (i2) edge [thick, dashed, ->, bend right = 10, draw=yafcolor1!60] (i7);
\draw (i3) edge [thick, ->, bend left = 10,  draw=yafcolor1!60] (i4);
\draw (i3) edge [thick, dashed, ->, bend right = 5,  draw=yafcolor1!60] (i6);
\draw (i5) edge [thick, dashed, ->, bend right = 10, draw=yafcolor1!60] (i4);
\draw (i5) edge [thick, dashed, ->, bend right = 10, draw=yafcolor7!60] (i7);
\draw (i5) edge [thick, ->, bend left = 10,  draw=yafcolor1!60] (i6);

\draw (11.5,-0.5) 
	edge [thick,->,draw=yafcolor1!60]  
	node [text=yafcolor1!70,above] {relay} 
	(12.5,-0.5);

\draw (11.5,-1) 
	edge [thick,->,draw=yafcolor2!60]  
	node [text=yafcolor2!70,above] {reply} 
	(12.5,-1);

\draw (11.5,-1.5) 
	edge [thick,->,draw=yafcolor7!60]  
	node [text=yafcolor7!70,above] {broadcast} 
	(12.5,-1.5);	

\node[text=black!70, font=\large] at (10.75,-2) {(b)};

\end{tikzpicture}
 \caption{\label{fig:toy_example}
A toy example showing the email communication network within a company. 
(a) The interaction network. 
Each edge corresponds to one interaction/email, 
labeled as ({\tt interaction id}, {\tt message topic}, {\tt timestamp}). 
(b) The corresponding \metagraph. 
Topics in both graphs are depicted using different colors. 
Edges in the \metagraph are depicted by a different color
according to their type (relay, reply, or broadcast). 
Edges are solid if they have small weight (topic dissimilarity). Otherwise, they are dashed. 
}
\end{figure}
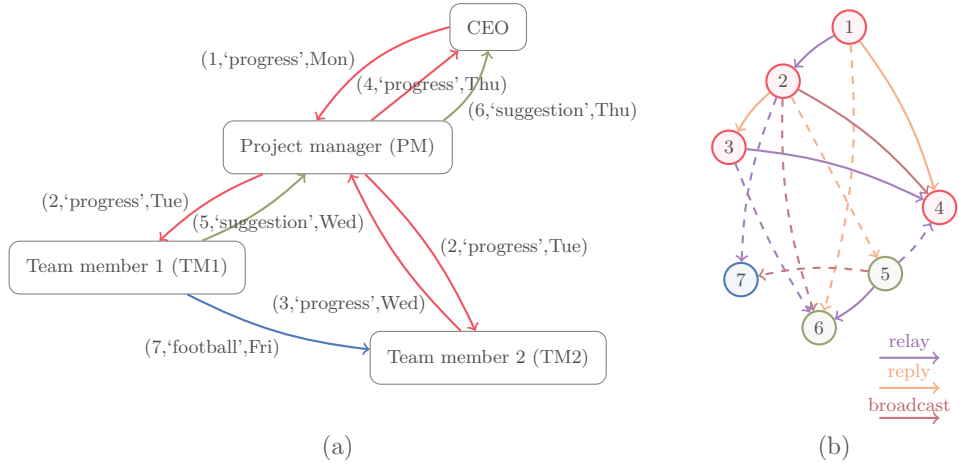

Motivated by the previous discussion, 
we formalize the task of interaction-network summarization as the problem of finding top-$\nevents$ trees 
in the transformed \metagraph $\mgraph=(\mnodes, \medges)$. 
We decompose this task into two sub-problems. 
First, we find a set of independent candidate events 
that are temporally and topically coherent. 
Since our goal is to summarize the interaction network
we aim to find large events. 
We show that this problem is the budget version of prize-collecting Steiner-tree problem in directed acyclic graphs. 
We provide three algorithms, among which a greedy approach performs the best.

The second sub-problem is to select $\nevents$ events that maximize the overall node coverage. 
This task maps to the maximum set-cover problem, and 
it can be approximated using a standard greedy algorithm. 
To speed up further our algorithm, 
we also propose a search strategy 
that avoids evaluating candidate events at all possible tree roots, 
but heuristically selects the most promising ones. 

\begin{example}
Consider the email communication network of a company, 
such as the one shown in Fig.~\ref{fig:toy_example}.
The interaction network is shown in Fig.~\ref{fig:toy_example}(a)
and the corresponding \metagraph in Fig.~\ref{fig:toy_example}(b).
The edges between interactions 
$(2, 4)$, $(1, 2)$, $(2, 3)$ in Fig.~\ref{fig:toy_example}(b) 
are examples of edge types broadcast, relay, and reply, respectively. 
In this toy example there are two main events. 
($i$) 
\textit{progress}: 
The {\sc ceo} asks a project manager ({\sc pm}) about progress on a project, 
and the {\sc pm} forwards the request to team members 1 ({\sc tm1}) and 2 ({\sc tm2}).
Later, {\sc tm1} reports back to {\sc pm}, who in turn reports back to {\sc ceo}.
The information flow of this event follows the interactions 
$1 \rightarrow 2 \rightarrow 3 \rightarrow 4$.
($ii$) \textit{suggestion}: 
Motivated by the first event, 
{\sc tm2} comes up with some suggestion, which she sends to {\sc pm}. 
The {\sc pm} finds the suggestion useful and forwards it to {\sc ceo}.
The information flow of this second event is $5 \rightarrow 6$. 
A third event, \textit{football}, is smaller in size, 
and it is not included in the top-2 events. 
Note that due to time ordering of the interactions, the \metagraph \mgraph is a directed acyclic graph.
\end{example}

The problem considered in this paper 
has many applications in different domains.
In our experimental evaluation, 
we focus on analyzing textual data in social media. 
We experiment with one email dataset (Enron) and three Twitter datasets. 
We provide a comparison of the different approaches, 
as well as many examples in which our methods discover meaningful events.

The contributions of this paper are summarized as follows\footnote{All scripts are available at \url{https://github.com/xiaohan2012/lst}}: 
\begin{itemize}
\item 
We propose a novel formulation for the problem of
discovering events that are temporally and topically coherent
in interaction networks, such as, online communication networks.
\item 
We present a transformation of the interaction network
to an \metagraph, 
which captures temporal and topical association of interactions
as well as the information flow in the network. 
This transformation helps to provide a cleaner abstraction
to the event-detection problem.
\item 
For the
problem of finding high-volume events
while satisfying constraints of temporal and topical coherence
we present and we evaluate three different algorithms:
a greedy approach, 
a dynamic-programming algorithm, 
and an adaptation to an existing approximation algorithm. 
\item 
We address the problem of finding 
the top-$\nevents$ events that summarize the network activity. 
The classic greedy algorithm is the standard way to approach this problem, 
but here, to speed-up the computations, we also propose and evaluate a search strategy 
that avoids construction of candidate events at all possible tree roots, 
but adaptively selects the most promising ones. 
\item 
We compare and analyze our algorithms on both synthetic and real datasets, 
such as Twitter and email communication. 
We show that our methods are able to detect meaningful temporal events.
\end{itemize}
 
\section{Related work}

\spara{Phrase-based event detection.}
The problem of detecting events in social media has attracted significant  attention.
Leskovec et al.~\cite{leskovec2009meme} and Yang et al.~\cite{yang2011patterns} 
treat events as short, distinctive phrases 
that propagate relatively intact through in a network.
Their work offers a graph formulation 
for clustering variants of phrases based on string edit distance.
Although their objective is similar to ours, there are significant differences. 
First, our methods focus on interaction networks, 
aiming to capture information flows in communication networks, rather than action networks. 
Second, we explicitly impose topic-coherence constraints, 
where the edit distance is insufficient for this goal. 
Third, instead of representing events by phrases, 
we derive higher-level representation using topic terms.

\spara{Text summarization.}
Text summarization techniques attempt to select a subset of sentences~\cite{celikyilmaz2011discovery} 
or tweets~\cite{inouye2011comparing} to summarize textual content. 
Similarly, we select a subset of interactions under a topic-coherence constraint. 
However, we also impose temporal coherence constraint, 
whereas they take a static view.

\spara{Statistical methods.}
Statistical and machine learning approaches for event detection are gaining increasing attention 
in recent years. Mathioudakis et al.~\cite{mathioudakis2010Twittermonitor} develop
an interactive system for identifying trends (events). 
The system first identifies ``bursty'' keywords, 
then clusters them based on co-occurrence and later 
performs trend analysis using dimension-reduction methods. 
Becker et al.~\cite{becker2011beyond} focus on online event identification. 
Their approach relies on online clustering techniques 
in order to discover topically-related tweets as an event and 
feature-based modeling in order to distinguish events from non-events. 
The difference of this approach with our work is that we offer a graph-theoretic
formulation. 

\spara{Graph-based methods.}
Other event-detection methods are based on constructing 
a word graph~\cite{weng2011event, meladianos2015degeneracy, cataldi2010emerging}. 
Weng et al.~\cite{weng2011event}  combines wavelet analysis and 
graph-partitioning techniques to cluster the words into events. 
Meladianos et al.~\cite{meladianos2015degeneracy} construct a word graph to represent a sequence of tweets, 
however, they focus on identifying key sub-events inside the sequence. 
Cataldi et al.~\cite{cataldi2010emerging} detect events by 
locating strongly connected components. 
Compared to those approaches, 
in this paper we explicitly model interactions, 
and take into account temporal constraints and topical-coherence constraints.  \section{Model}
\label{sec:model}

An \emph{interaction network} $\graph = (\nodes, \edges)$ 
consists of a set of $\nnodes$ nodes~$\nodes$ and a set of \nedges time-stamped interactions~\edges 
between pairs of nodes. 
\edges is represented as: 
\[
\edges = \set{(u_i, v_i, \info_i, t_i)}, 
\text{ with } i = 1,\ldots, \nedges,
\text{ such that }  u_i, v_i \in \nodes,~t_i \in \real,~ \info_i\in\real^\ntopics, 
\]
indicating that nodes $u_i$ and $v_i$ interacted at time $t_i$. 
Each interaction is annotated with textual content represented by $\info_i$. 
The representation is independent to our main methodology. 
We can use various text modeling techniques such as bag-of-words representation or latent Dirichlet allocation 
(LDA)~\cite{blei2003latent}. 

For generality we consider that interactions are directed. 
More than one interaction may take place between a pair of nodes, with different timestamps. 
Conversely, more than one interaction may take place at the same time, between different nodes.  
Online communication networks, 
such as email networks, 
are examples of interaction networks.

Given an interaction network \graph 
we construct a directed weighted \emph{\metagraph} 
$\mgraph=(\mnodes, \medges, \mweight)$. 
The vertices $\mnodes$ in $\mgraph$ correspond to the interactions \edges in \graph. 
There is an edge from vertex 
$\none=(u_i, v_i, \info_i, t_i)\in\mnodes$ to a vertex
$\ntwo=(u_j, v_j, \info_j, t_j)\in\mnodes$ if the following holds:

\begin{enumerate}
	\item{Interaction $\none$ takes place before interaction \ntwo (time comprehension): $t_i\leq t_j$.}
	\item{Information comprehension takes place in one of the following ways: 
	\begin{enumerate}
		\item{interactions \none and \ntwo share the same start node in \nodes: $u_i=u_j$ (broadcast);}
		\item{the end node of interaction \none is the start node of interaction \ntwo and the end node of \ntwo is not the start node of \none: $v_i=u_j$ and $v_j\not=u_i$ (relay);}
		\item{the end node of an interaction \none is the start node of an interaction \ntwo and the end node of \ntwo is the start node of \none: $v_i=u_j$ and $v_j=u_i$ (reply)}.
	\end{enumerate}}
\end{enumerate}
Note, that due time comprehension the \mgraph is a directed acyclic graph (DAG).

For the edges of the \metagraph \mgraph 
we use weights to measure the topical (dis)similarity
between interactions. 
Thus, 
given two interactions 
$(u_i, v_i, \info_i, t_i)$ and
$(u_j, v_j, \info_j, t_j)$
connected by an edge in \mgraph,
our edge-weighting function  
$c:\medges\rightarrow\real$ is a distance function between topic vectors $\info_i$ and $\info_j$.

Finally, given a meta-graph 
$\mgraph=(\mnodes, \medges, \mweight)$ and a time interval $\twindow$ we define the \emph{time-induced meta-graph}  
$\mgraph(\twindow) = (\mnodes(\twindow), \medges, \mweight)$, where $\mnodes(\twindow)$ are the interactions that occur in $\twindow$: $\mnodes(\twindow) = \set{(u, v, \info,t) \in \mnodes \mid s \leq t \leq f}.$

 \section{Problem formulation}
\label{sec:problems}

We aim at summarizing the top-$\nevents$ events in an interaction network. 
We define an \emph{event} to be a rooted subtree \event of the \metagraph \mgraph. 
An event naturally has a source vertex (or interaction) and is spread in the network.
We are interested in events of high volume, 
which translates into a large number of iterations included into the tree \event.
We are also interested in events with temporally close and topically coherent interactions. 

These aspects can be incorporated into the optimization cost function in different ways. 
Our primary objective is to obtain $\nevents$ events that have high enough coverage to represent the whole network, 
and thus, we aim to maximize the number of interactions that are included in the event. 
To incorporate temporal and topical coherence we set constraints 
on the time interval spanned by the event tree (temporal coherence),  
and on total weight of its edges (topical coherence). 

To simplify the problem of finding the best $\nevents$ events, 
we decompose the main task into two subproblems: 
(1) finding a set of independent candidate events 
that satisfy the constraints and maximize volume of interactions, and 
(2) selecting the top-$\nevents$ events to maximize total coverage.
The first problem is defined as follows.

\begin{problem}
\label{prob:tree}
\textbf{Time-constrained maximum tree (\tpcst)}: 
Given an \metagraph $\mgraph=(\mnodes, \medges, \mweight)$, 
a root vertex $\treeroot \in \mnodes$, 
time budget \tbudget, and 
dissimilarity budget \distbudget, 
find a directed subtree $\event=(\enodes,\eedges)\subseteq\mgraph$, rooted at \treeroot, 
which satisfies the constraints 
\[ 
\sum_{e\in\eedges}\mweight(e)\leq\distbudget \quad\text{and}\quad 
(\max_{\none\in\enodes}t_\none - \min_{\ntwo\in\enodes}t_\ntwo)\leq\tbudget,\]
while maximizing the number of vertices $|\enodes|$.
\end{problem}

Note that the time constraint can be omitted, 
if we restrict the input graph to be induced by the time interval $[t_r,t_r+\tbudget]$, 
where $t_r$ is the root timestamp.
By omitting the time constraint, our problem can be written as follows. 

\begin{problem}
\label{prob:PCST}
\textbf{Maximum tree (\pcst)}: 
Given a weighted directed acyclic graph  
$\mgraph(\twindow)=(\mnodes(\twindow), \medges, \mweight)$, a root vertex $r$,  
and cost budget \distbudget, 
find a subtree $\event=(\enodes,\eedges)\subseteq\mgraph(\twindow)$, 
rooted at $r$, that satisfies $\sum_{e\in\eedges}\mweight(e)\leq\distbudget$
while maximizing the number of vertices $|\enodes|$. 
\end{problem}

We observe that \pcst is directly related to 
\textit{budget} version of the prize-collecting Steiner-tree problem (PCST)~\cite{johnson2000prize}. 
However, 
we are dealing with a special case of the budget PCST, 
as vertex prize is uniform and our input graph is a DAG.
Despite so, this special case is still \np-hard. 

\begin{proposition}
\pcst is \np-hard.\footnote{Proof can be found at appendix. } 
\end{proposition}

As the interaction network is likely to contain more than one event, 
we are interested in finding $k$ events that describe different aspects of the whole network while covering as much
activity as possible.
This is captured in the following problem formulation.
\begin{problem}
\label{prob:ktrees}
\textbf{Maximum $\nevents$ trees (\ktrees)}: 
We are given an \metagraph $\mgraph=(\mnodes, \medges, \mweight)$ and $k\in\naturals$. 
Find a set of $k$ vertex-disjoint trees $\events=\{\event_1,\dots\event_k\}$, with each event tree $\event=(\enodes,\eedges)\in \events$ to be a subgraph of \mgraph rooted in some $r_i\in \mnodes$, such that the total number of spanned interactions $|\cup_{\event=(\enodes, \eedges)\in \events} \enodes|$ is maximized.
\end{problem}

It is easy to observe that this problem is equivalent to maximum $k$-coverage problem and thus is \np-hard.
To solve \ktrees efficiently, 
we consider the question of sampling as few root vertices as possible 
so that the major events can still be captured. Real-world networks consist of millions of interactions so it is 
impractical to calculate candidate event trees rooted at each vertex.

 \section{Algorithms}
\label{sec:algo}

\subsection{Approximating \pcst}

For finding the best tree, 
as defined by \pcst, 
we consider three algorithms.
Recall that for \pcst
we are working with the \metagraph \mgraph, 
and that a root vertex is fixed. 

\spara{Greedy tree growing:}
The greedy algorithm starts from the root and
builds the event tree by adding one vertex (interaction) at a time. 
At each step the algorithm selects the edge with the minimum cost (topic dissimilarity) 
from the cutset of the current tree. 
This choice aims to maximize the topical coherent of the event discovered. 
The running time is $\bigO(\abs{\mnodes}^2)$.

\spara{Directed Steiner tree algorithm (DST):}
Recall that \pcst corresponds to the \textit{budget} PCST problem. 
Our second algorithm is inspired by an approach proposed by Johnson et al.~\cite{johnson2000prize}, 
where the the \textit{budget} PCST problem can be solved by the \textit{quota} PCST problem
using binary search. 
In our case, the prizes of all vertices are uniform, 
thus the \textit{quota} PCST problem is equivalent to \textit{$k$-minimum spanning tree}. 
The latter problem can be solved by an algorithm for finding directed Steiner trees (DST),
such as the algorithm proposed by Charikar et al.~\cite{charikar1999approximation}.
Thus, our second algorithm uses the DST algorithm, within a binary search to find 
an event that satisfies the budget constraint. 
The DST algorithm takes four arguments, 
$\mgraph,\, \treeroot,\, \charikarterminals$, and $\charikarlevel$, 
where $\charikarterminals$ is a set of terminal nodes and $\charikarlevel$ is a 
parameter that provides a quality-of-approximation vs.\ efficiency trade-off. 
The running time of the algorithm is 
$\bigO(\abs{\mnodes}^{\charikarlevel} \abs{\charikarterminals}^{2\charikarlevel})$. 
In our case, $\charikarterminals=\mnodes$, thus 
the running time is $\bigO(\abs{\mnodes}^{3\charikarlevel})$.
We use $\charikarlevel=1$ but still the algorithm is mainly of theoretical interest 
and not practical for large datasets. 

\spara{Dynamic programming algorithm (DP):}
The third algorithm we present
is inspired by the idea that when the input DAG is a tree, 
the problem can be solved optimally using a simple dynamic programming approach. 
We investigate two approaches to adapt this algorithm for general (non-tree) DAGs.

In the first approach, we slightly modify the dynamic programming algorithm to make sure the result is a tree. Specifically, when attempting to connect the current node with the subtrees of its children, 
we enforce the condition that the subtrees cannot have any common nodes. 
In the second approach, we transform the input DAG into a tree and then apply 
the original dynamic programming algorithm. 
Specifically, we first calculate single-source shortest paths from $\treeroot$ to all vertices of \mgraph 
using Dijkstra's algorithm and then apply the dynamic programming algorithm. 
For integer edge weights and a tree input, the running time is $\bigO(\abs{\mnodes} \distbudget^2)$.
In our case, edge weights are real numbers, 
so we discretize the weights to some decimal digits.

\subsection{Approximating \ktrees}

Once we have computed a set of candidate event trees using any algorithm for \pcst, 
we need to select $\nevents$ event trees from the candidate set so that vertex coverage is maximized. 
This is essentially the \textit{maximum coverage} problem. 
A standard greedy algorithm gives approximation ratio 
$(1 - \frac{1}{e})$ in time $\bigO(\abs{\mnodes}^2)$~\cite{vazirani2013approximation}.

\subsection{Root sampling strategy}

One issue with the greedy max-cover algorithm discussed above, 
is that all candidate root vertices need to be tested 
before selecting the one that greedily optimizes the coverage. 
This is an expensive computational task. 
To speed up the algorithm for finding top-\nevents trees,  
we propose a simple root-sampling strategy that ranks roots according 
to their potential of maximizing \pcst. 
   
For every sampled root $r$ we construct a candidate tree \rooteddag and evaluate
\textit{event size upper bound} $\eupperbound(\rooteddag, \distbudget)$ of DAG $\rooteddag$ with budget $B$, defined as:
\[
\eupperbound(\rooteddag, \distbudget) = \max\limits_{\forest' \in \forest(\rooteddag)} 
\set{\abs{\forest'.\mnodes} \mbox { such that } \sum\limits_{e \in \forest'.\medges} \mweight(e) \le \distbudget}
\]
where $\forest(\rooteddag)$ is a set of all forests containing $\rooteddag.\treeroot$ (the root of DAG $\rooteddag$).

It is easy to see the optimal tree 
$\event(\rooteddag, \treeroot, \distbudget)$ 
cannot have size greater than $\eupperbound(\rooteddag, \distbudget)$, thus  $\eupperbound(\rooteddag, \distbudget)$ is indeed an upper bound.

Define the \textit{minimum in-edge} of a vertex $\node$ as 
\[
e^{*}(\mgraph, \node) = \argmin\limits_{e' \in \inedges(\mgraph,\node)} \mweight(e'),\] 
where $ \inedges(\mgraph,\node) = \set{e \in \mgraph.\medges \mid e.\none = \node}$. 
$\eupperbound$ can be computed efficiently as follows. Consider only nodes, which belong to $[t_r,t_r+\tbudget]$ time interval, 
where $t_r$ is the root timestamp. Start constructing an event \rooteddag by adding root $r$ and its child with the lightest edge. Now sort all other nodes by cost of their minimum in-edge cost in increasing order; greedily add nodes with their minimum in-edge to the event \rooteddag and stop when budget constraint \distbudget is reached. \eupperbound is a number of nodes in the event \rooteddag. Note that \rooteddag is a forest, as we do not care about connectivity during construction.

Our root sampling strategy first ranks all the vertices by \eupperbound. 
Then it sequentially selects vertices from the ranked list.

 \section{Experimental evaluation}
\label{sec:exp}

As no datasets with ground-truth events are available to us, we validate our approach by using synthetic datasets and by case studies.
For the experiments with synthetic datasets:
(1) we plant events (considered as ground truth) within random interaction networks; 
(2) we then apply our algorithm to find events in those synthetic data; 
(3) we measure the precision and recall of the discovered events with respect to ground-truth.
For the case studies we apply our algorithm on \enron and on Twitter data, then examine the events we discover, and map them on real known historical events based on textual content and time period.
As means of exploratory data analysis, we also visualize the event trees in order to show the information flow within the event.

\subsection{Datasets and preprocessing}

\spara{Synthetic data.}
We generate synthetic datasets in two steps:
  (1) we generate ground-truth event trees; 
  (2) we inject noise interactions. 
Each event is generated independently using the model by Kumar et al.~\cite{kumar2010dynamics}, 
which constructs a tree by iteratively adding random edges. 
We sample a sender, recipients, timestamp and a topic vector randomly for each node. 

\begin{table}[t]
\centering
\caption{Network statistics on real datasets. 
    Singleton interactions in the \metagraph are removed. }
\begin{tabular}{c|cc|ccc}
\toprule
\textbf{Datasets} &  \multicolumn{2}{c}{\textbf{Interaction networks}} &  \multicolumn{3}{c}{\textbf{Interaction meta-graphs}}  \\
{} & \#nodes & \#edges & \#nodes & \#edges &    Period \\
\midrule
\enron     &      1144 &      2106 &        812 &      21297 & 1998-10-30 - 2002-02-13          \\
\beefban   &     11895 &     33584 &      26317 &      75870 & 2015-03-03 - 2015-03-05 \\
\ukraine   &     16218 &     59096 &      46540 &     142746 & 2015-02-27 - 2015-03-03  \\
\baltimore &     38541 &    102139 &      61501 &     132012 & 2015-04-26 - 2015-04-28 \\
\bottomrule
\end{tabular}
\label{tbl:dataset_stat}
\end{table}

\spara{Real-world data.}
We use two real-world datasets: 
email (\enron) and Twitter. Dataset statistics are given in Table~\ref{tbl:dataset_stat}. 
\textbf{\enron:} we use a preprocessed version of the original Enron dataset~\cite{fiore2004uc}. 
\textbf{Twitter:} we use Twitter datasets extracted for three hashtags,
each one containing a specific hashtag. 
The hashtags are \beefban, \baltimore and \ukraine. There is a interaction from a user $u$ to a user $v$, if the tweet of user $v$ contains username of $u$.
The Twitter datasets are provided by Garimella et al.~\cite{garimella2015quantifying}. 

\spara{Preprocessing.}
We observe the phenomenon that the same person sends the same   (or very similar) messages multiple times, 
especially on Twitter. 
Our methods are easily misled by the sheer amount of redundant messages. 
To avoid this problem, we merge similar messages from the same sender into one. 
We consider two messages \textit{similar} if 
  (1) they are sent by the same user, 
  (2) their Levenshtein edit distance ratio is below 10\%, 
  (3) their time distance is relatively small  (e.g., one day). 
In the newly-merged message, the text content, timestamps are copied from the earliest message. 
Recipients are the union of all recipients. 

We take different approaches for representing interaction content in \enron and Twitter. 
For \enron, we train a topic model using \textit{gensim} \footnote{https://radimrehurek.com/gensim/models/ldamodel.html}.
We assign each interaction a topic vector and use cosine distance to compute edge weight. 

Measuring tweet similarity is an open challenge due to its short length and conciseness. 
We took an ensemble approach where vector representation comes from several models. 
Besides topic vectors, we use also  ($i$) bag-of-word (BoW) with tf-idf re-weighting and 
 ($ii$) hashtags included in each tweet. 
For BoW and hashtag representations, we use cosine and Jarccard distance for weight assignment, respectively. 
Last, we sum up the three distances. 
For topic modeling, for both \enron and Twitter datasets, 
we use 10 topics, batch size 100 and run it for 10 iterations.

\subsection{Results on synthetic datasets}

We evaluate five different algorithms for finding the best event:
 (1) greedy tree growing  (\greedy),
 (2) binary search using Charikar's DSP algorithm  (\aquota),
 (3) dynamic programming without preprocessing  (\lst),
 (4) dynamic programming with Dijkstra preprocessing  (\lstdij), and
 (5) random tree growing  (\random) as a baseline.
The \random algorithm mimics the \greedy, but it selects a random edge to grow at each step. 
We compare quality of solutions obtained on datasets with various noise level.
We define \textit{noise level} as a number of noise interactions divided by the total number of interactions of all events. 
For the DSP algorithm we set level parameter $\charikarlevel=1$, 
as we have insufficient memory for experiments with larger values. 

\spara{Different noise levels.}
To compare the capability of the algorithms to find \textit{one} best event, 
we generate a sequence of datasets with increasing noise levels and only one event of size 20 
 (containing 20 nodes). 
We set ground-truth values of $\tbudget, \distbudget, \treeroot$ for parameters in \pcst.
We consider three types of measurements:  (1) precision, recall, and F1, 
 (2) the value of objective function, 
and  (3) the running time. 
Log scale is applied in the case of running time as difference between algorithms is of magnitudes order.

\begin{figure}[t]

\begin{center}
\begin{minipage}{0.7\linewidth}
\centering
\subfloat[]{
  \includegraphics[width=1.0\textwidth]{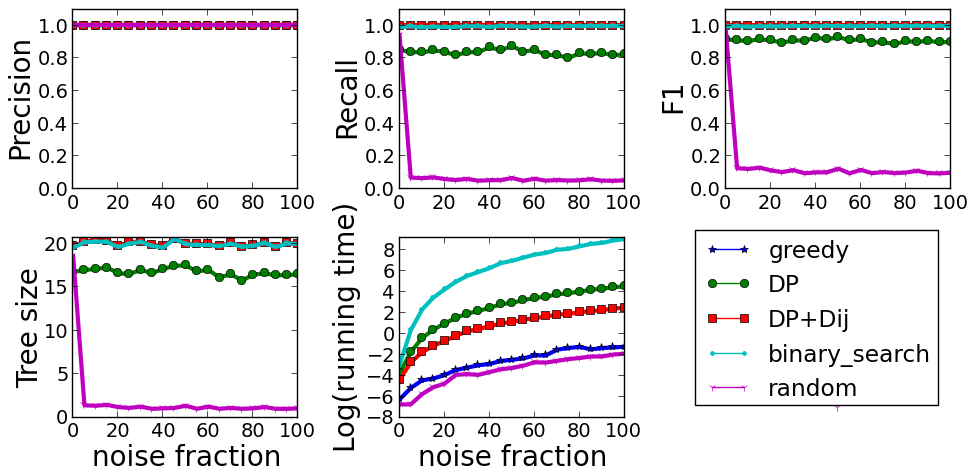}
  \label{main:a}
}
\end{minipage}\end{center}
\begin{center}
\begin{minipage}{0.7\linewidth}
\centering
\subfloat[]{
  \includegraphics[width=1.0\textwidth]{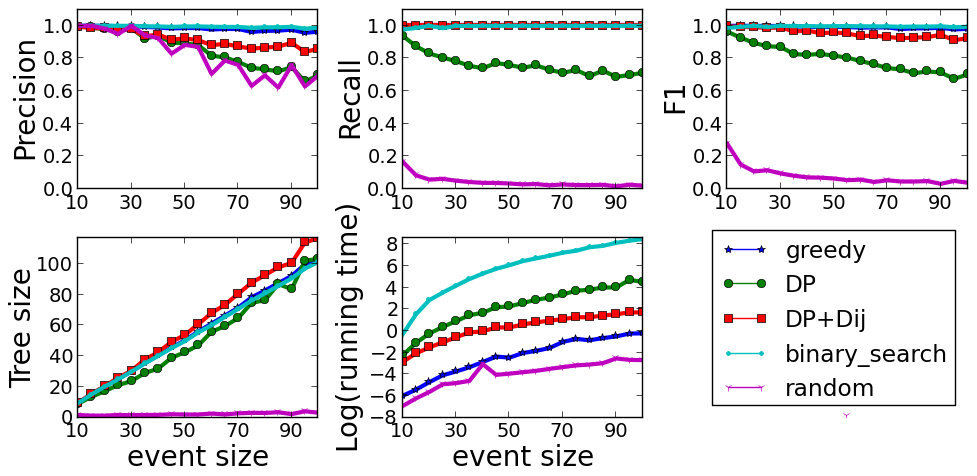}
  \label{main:b}
}
\end{minipage}
\end{center}
\caption{ (a) Performance of the algorithms under noise levels from 0 to 100 with step size 0.5. 
Results are averaged over 50 repetitions. 
 (b) Performance of the algorithms on synthetic dataset with noise level 20 and varying event size from 10 to 100 at step size 10. Measurement values are averaged from 50 rounds.}
\label{fig:synthetic_exp}
\end{figure}

In Fig.~\ref{fig:synthetic_exp} (a),
we see that all our algorithms outperform the trivial \random baseline. 
Although \greedy is a simple heuristic, 
its performance is among the top. 
Dijkstra preprocessing for \lst improves both F1 and computational time. In the contrary,
\aquota consumes much time, even though it is among the best in other measurements. 
Notice that \random achieves high precision because it can select a wrong edge that violates the budget constraint at the first few steps and terminate.

\spara{Different event sizes.}
We also study how the algorithms perform in extracting events of different sizes. 
The experiment setting is similar to the above, but the noise level is fixed to 20,
while the event size varies. 
In Fig.~\ref{fig:synthetic_exp} (b), \greedy, \aquota are among the best in terms of precision, recall, F1 and set cover objective, 
whereas \lstdij is slightly worse due to needed edge weight discretization. 
Again, preprocessing for \lst improves performance. 
Running time comparison is consistent with the previous case.

\subsection{Parameter effects on real datasets}

\begin{figure}[t]
  \centering
  \includegraphics[width=\textwidth]{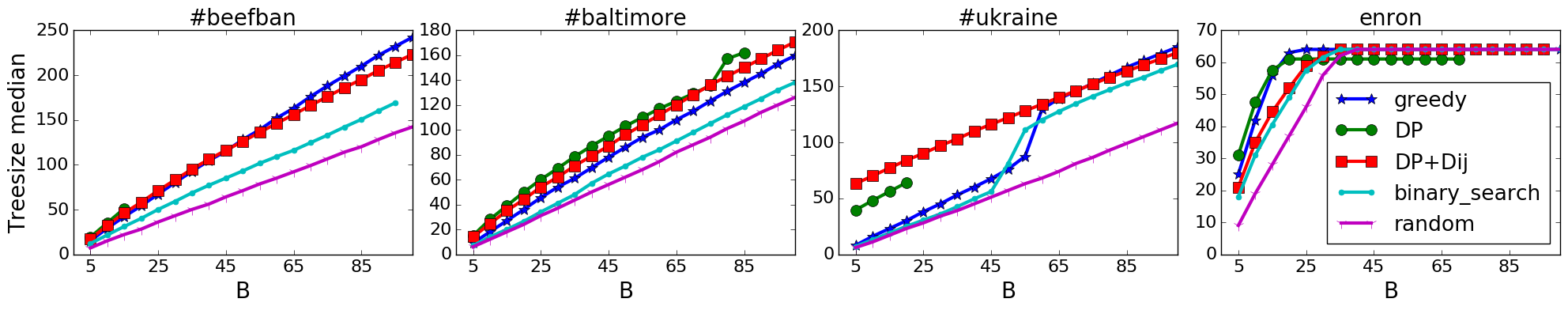}
  \caption{Effect of $\distbudget$ on the median of tree sizes for different datasets. Note that for \ukraine and $\distbudget>25$, the DP algorithm fails to complete the experiments as it consumes excessive amount of memory.}
  \label{fig:budget_exp}
\end{figure}

\spara{Effect of $\distbudget$.}
We evaluate the effect of topic dissimilarity budget $\distbudget$
on the tree size objective in \pcst. 
We randomly sample 100 roots for each dataset. 
$\distbudget$ varies from 0 to 100 at a step size of 5.0. 
For Twitter and \enron dataset, $\tbudget$ is set to 1 day and 4 weeks respectively. 
We take the median of all trees returned by each algorithm  (Fig.~\ref{fig:budget_exp}). 

In \enron, we observe a converging effect on both objectives as the dataset is relatively small,  
while this is not the case in all Twitter datasets. 
In practice, \greedy is the best performing algorithm, 
as it is both competitive in maximizing the objective function and it is computationally efficient.

\begin{figure}[t]
	\centering
	\includegraphics[width=\textwidth]{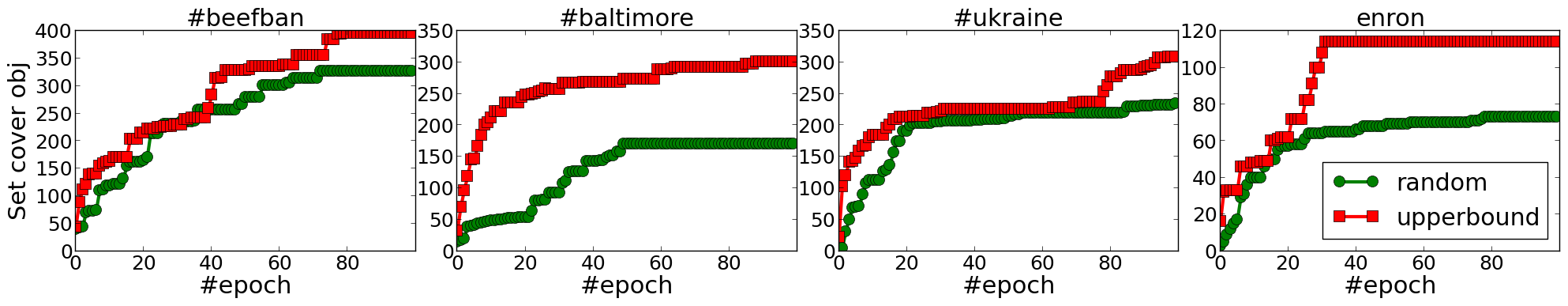}
	\caption{Performance of Different sampling schemes on real datasets: $\nevents=10$. For Twitter, $\distbudget=15.0, \tbudget=\text{1 day}$. For \enron, $\distbudget=10.0, \tbudget=\text{4 weeks}$. 100 unique roots are selected based on the sampling scheme.}
	\label{fig:sampler_experiment}
\end{figure}

\spara{Sampling scheme comparison.}
We compare two sampling schemes in real data setting: 
 (1) random root sampling  (\random) as the baseline, 
 (2) ranking roots by event size upperbound  (\upperbound)~\ref{sec:algo}.
For each scheme, the set cover objective is recorded whenever a new candidate is added. 
As we can see in Fig.~\ref{fig:sampler_experiment}, the event size upper-bound heuristic helps
to discover better solutions,  
especially for \baltimore and \enron.

\begin{figure}[t]

\begin{center}
\begin{minipage}{.35\linewidth}
\centering
\subfloat[]{
  \label{main:a}
  \def\svgwidth{\columnwidth}
  \begingroup  \makeatletter  \providecommand\color[2][]{    \errmessage{(Inkscape) Color is used for the text in Inkscape, but the package 'color.sty' is not loaded}    \renewcommand\color[2][]{}  }  \providecommand\transparent[1]{    \errmessage{(Inkscape) Transparency is used (non-zero) for the text in Inkscape, but the package 'transparent.sty' is not loaded}    \renewcommand\transparent[1]{}  }  \providecommand\rotatebox[2]{#2}  \ifx\svgwidth\undefined    \setlength{\unitlength}{271.63001091bp}    \ifx\svgscale\undefined      \relax    \else      \setlength{\unitlength}{\unitlength * \real{\svgscale}}    \fi  \else    \setlength{\unitlength}{\svgwidth}  \fi  \global\let\svgwidth\undefined  \global\let\svgscale\undefined  \makeatother  \begin{picture}(1,0.94889596)    \put(0,0){\includegraphics[width=\unitlength]{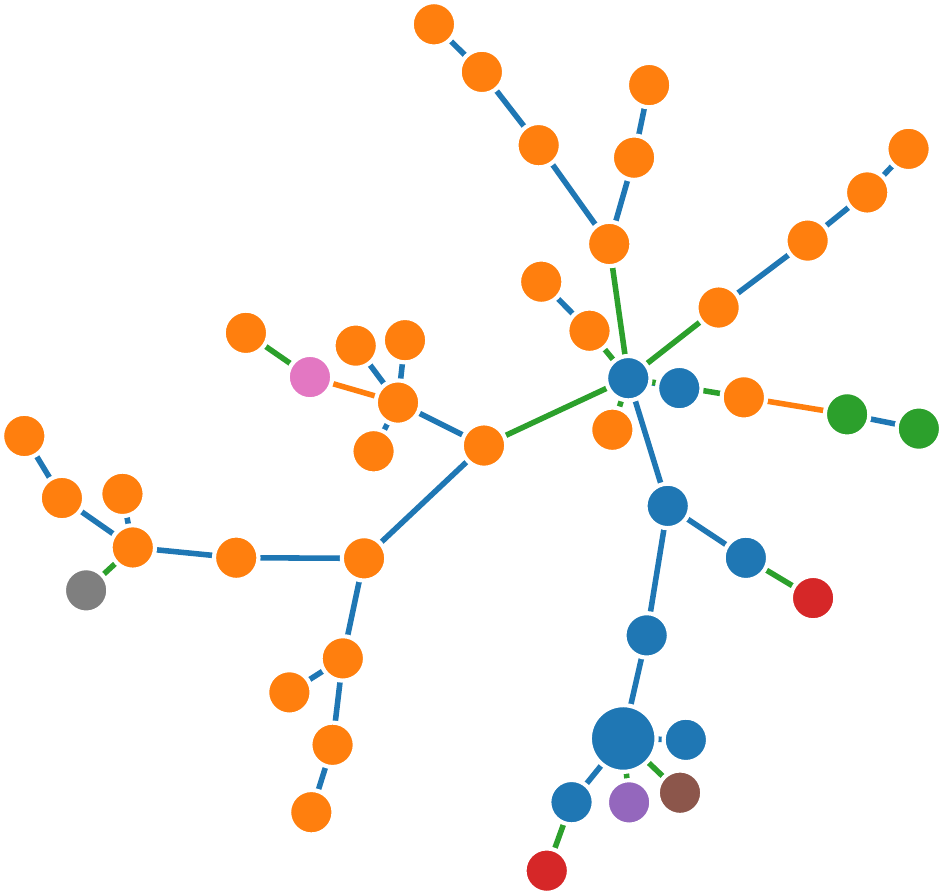}}  \end{picture}\endgroup }
\end{minipage}\hspace{.1\linewidth}
\begin{minipage}{.35\linewidth}
\centering
\subfloat[]{
  \label{main:b}
  \def\svgwidth{\columnwidth}
  \begingroup  \makeatletter  \providecommand\color[2][]{    \errmessage{(Inkscape) Color is used for the text in Inkscape, but the package 'color.sty' is not loaded}    \renewcommand\color[2][]{}  }  \providecommand\transparent[1]{    \errmessage{(Inkscape) Transparency is used (non-zero) for the text in Inkscape, but the package 'transparent.sty' is not loaded}    \renewcommand\transparent[1]{}  }  \providecommand\rotatebox[2]{#2}  \ifx\svgwidth\undefined    \setlength{\unitlength}{292.66806669bp}    \ifx\svgscale\undefined      \relax    \else      \setlength{\unitlength}{\unitlength * \real{\svgscale}}    \fi  \else    \setlength{\unitlength}{\svgwidth}  \fi  \global\let\svgwidth\undefined  \global\let\svgscale\undefined  \makeatother  \begin{picture}(1,0.87909742)    \put(0,0){\includegraphics[width=\unitlength]{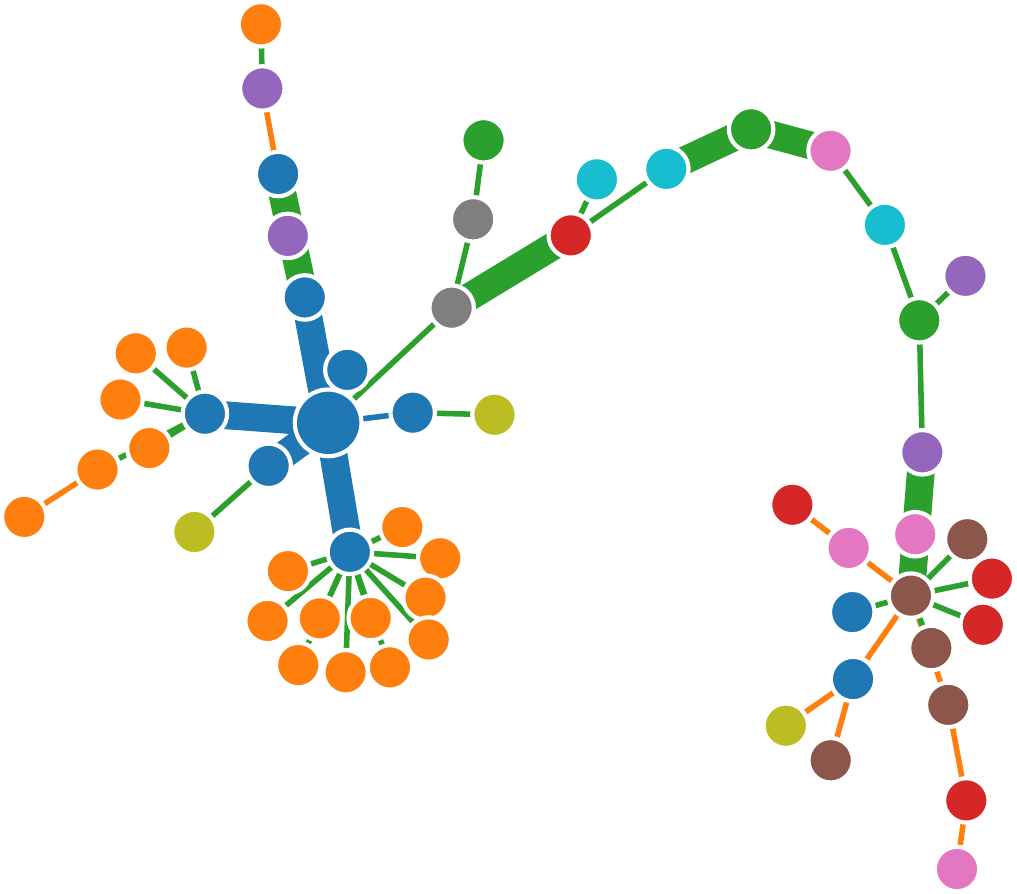}}  \end{picture}\endgroup }
\end{minipage}
\end{center}
\caption{Tree computed from \beefban given fixed root by \greedy (a) and \lstdij (b), which achieves tree size 46 and 57 respectively. Root, $\distbudget=30$ and $\tbudget=\text{1 day}$ are the same for both algorithms. Edges with weight $\ge$ 0.8 are wider. In the tree by \greedy, no edges with weight $\ge$ 0.8 are selected. Nodes are colored by senders and edges are colored by its type (broadcast: blue, relay: green, reply: orange)}
\label{fig:trees-by-algos}
\end{figure}

\spara{Event trees by different algorithms.}
We compare the behaviours of the algorithms for \pcst in real-world datasets. 
In Fig.~\ref{fig:trees-by-algos}, the trees are produced by \greedy, and \lstdij are given the same root and budget. 
The \greedy algorithm avoids to select heavy edges with weights larger than 0.8 due to its local search strategy whereas \lstdij achieves larger tree by selecting a few heavy edges. 
Therefore we expect  \greedy to produce more topically-coherent events as the pairwise dissimilarity between nodes tend to be smaller. 

\subsection{Case study in \enron dataset}

\begin{figure}[t]

\begin{minipage}{1.0\linewidth}
\centering
\subfloat[]{
  \includegraphics[width=1\textwidth]{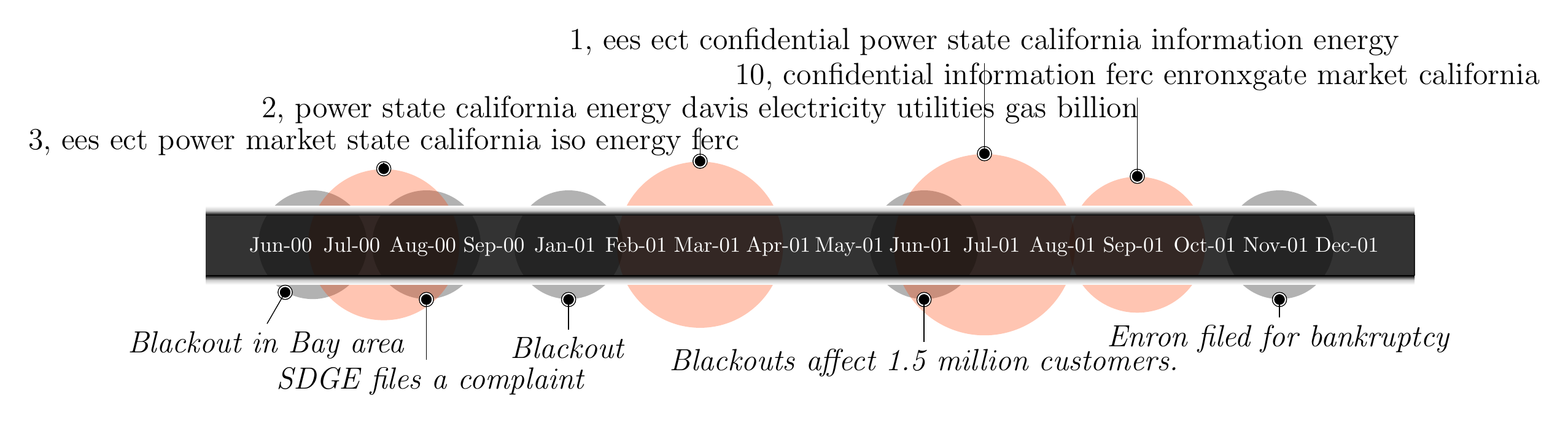}
  \label{main:a}
}
\end{minipage}
\begin{minipage}{1.0\linewidth}
\centering
\subfloat[]{
  \includegraphics[width=1\textwidth]{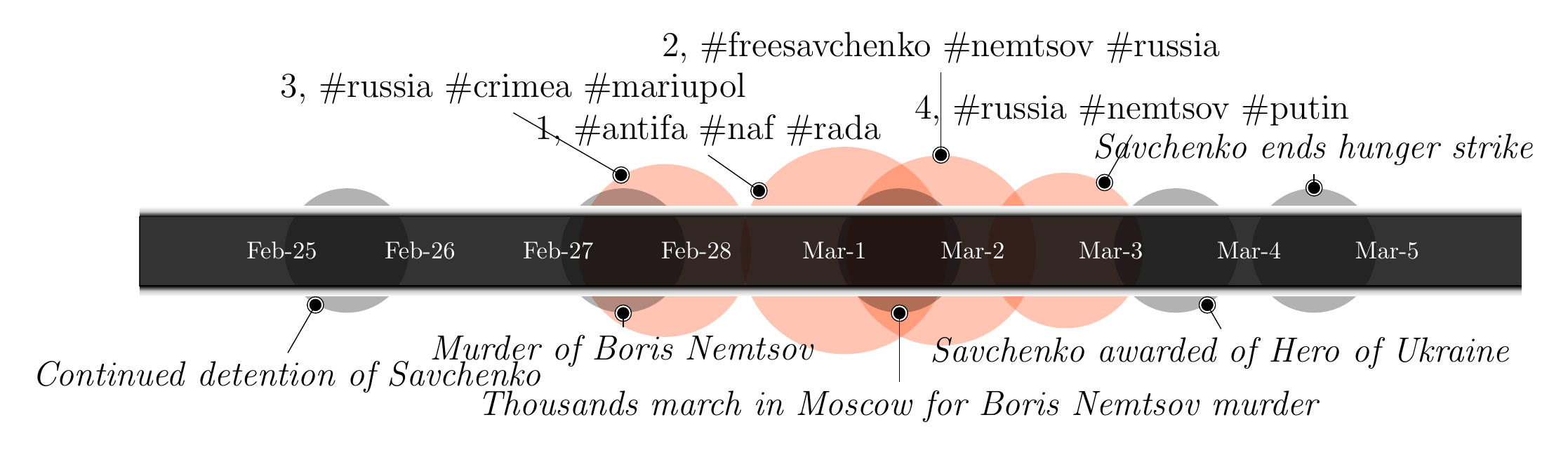}
  \label{main:b}
}
\end{minipage}
\caption{Timeline with extracted events (larger red circle) and publicly recognised events (smaller black circle and italic text) for Enron (a) and \ukraine (b).  ($a$) highlights events on Enron's energy scandal and bankruptcy. Event 3, 2, 1 and 10 are displayed. The larger the circle, the larger the event size is. For each event, top topic terms are displayed. 
In ($b$), top-4 events are displayed with the top hashtags. Event 2 and 4 maps to the murder of Boris Nemstsov (\#nemstsov), while event 4 also contains tweets on freeing Savchenko (\#freesavchenko). 
Event 1, 3 is about other related issues.
}
\label{fig:timeline}
\end{figure}

\begin{figure}[t]

\begin{minipage}{1.0\linewidth}
\centering
\subfloat[]{
  \includegraphics[height=0.12\textheight]{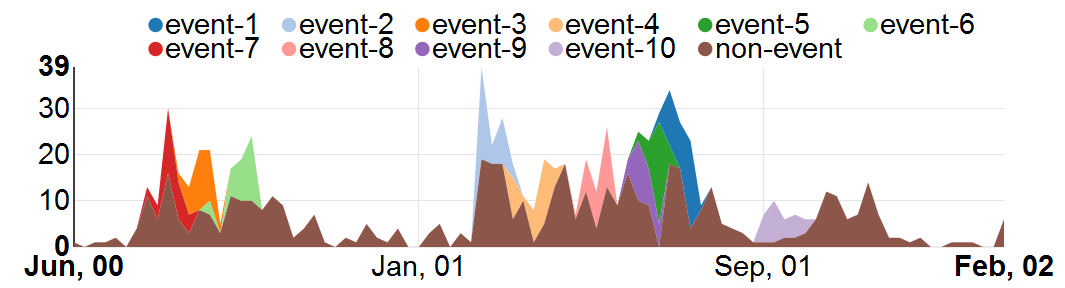}
  \label{main:a}
}
\end{minipage}
\begin{minipage}{1.0\linewidth}
\centering
\subfloat[]{
  \includegraphics[height=0.12\textheight]{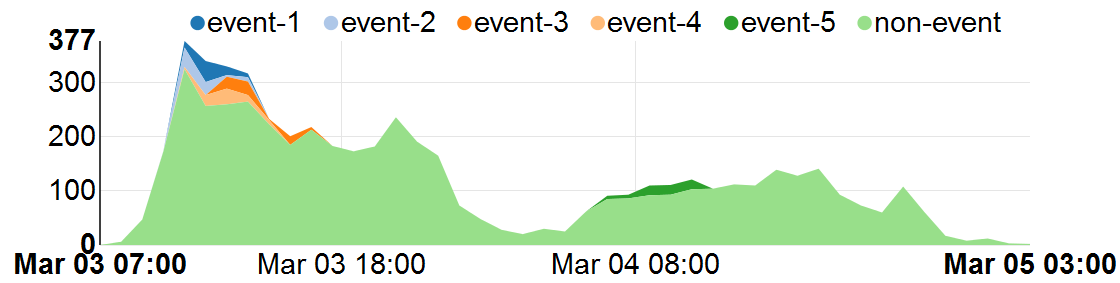}
  \label{main:b}
}
\end{minipage}
\caption{Stacked area graph of interaction frequency against time. \enron (a) contains top-10 events. \beefban (b) contains top-5 events.}
\label{fig:stream_graphs}
\end{figure}

We sample 50 nodes using \upperbound scheme and applied \greedy algorithm with $\distbudget=10, \tbudget=\text{28 days}$. 
First, we observed that the events can be grouped into two types: 
(1) California Energy Crisis,\footnote{https://en.wikipedia.org/wiki/California\_electricity\_crisis} 
(2) investigation into Enron's scandal.\footnote{https://en.wikipedia.org/wiki/Enron\_scandal}
In Fig.~\ref{fig:timeline} (a), we annotated the real world events about the crisis happening during the timespan of the dataset. 
We found shortly after each major blackout, there is at least one extracted events about it. 
And before Enron filed bankruptcy, Federal Energy Regulatory Commission (FERC) investigated Enron. 
Second, in Fig.~\ref{fig:stream_graphs} (a), extracted events tend to occur at the peak of the volume plot. 

\subsection{Case study in Twitter datasets}

We use the same parameters for all three datasets as they have similar size and timespan. Events are extracted by selecting 100 roots using \upperbound and using \greedy algorithm with $\distbudget=50$ and $\tbudget=\text{1 day}$.

\spara{\ukraine.}
Ukraine crisis
arouses media war on Ukraine and Russia.\footnote{
https://en.wikipedia.org/wiki/Ukrainian\_crisis.
}  
We observe some of the detected events align well topically and temporally with the actual events in Fig.~\ref{fig:timeline} (b). 
However, topics are mixed inside some other events. 
For example, topics on both \#nemstsov and \#freesavchenko 
 are detected in event 2. This is expected due to the local similarity measurement in \pcst.

\begin{figure}[t]

\begin{minipage}{.5\linewidth}
\centering
\subfloat[]{
  \def\svgwidth{\columnwidth}
  \begingroup  \makeatletter  \providecommand\color[2][]{    \errmessage{(Inkscape) Color is used for the text in Inkscape, but the package 'color.sty' is not loaded}    \renewcommand\color[2][]{}  }  \providecommand\transparent[1]{    \errmessage{(Inkscape) Transparency is used (non-zero) for the text in Inkscape, but the package 'transparent.sty' is not loaded}    \renewcommand\transparent[1]{}  }  \providecommand\rotatebox[2]{#2}  \ifx\svgwidth\undefined    \setlength{\unitlength}{512.57292331bp}    \ifx\svgscale\undefined      \relax    \else      \setlength{\unitlength}{\unitlength * \real{\svgscale}}    \fi  \else    \setlength{\unitlength}{\svgwidth}  \fi  \global\let\svgwidth\undefined  \global\let\svgscale\undefined  \makeatother  \begin{picture}(1,0.82515995)    \put(0,0){\includegraphics[width=\unitlength]{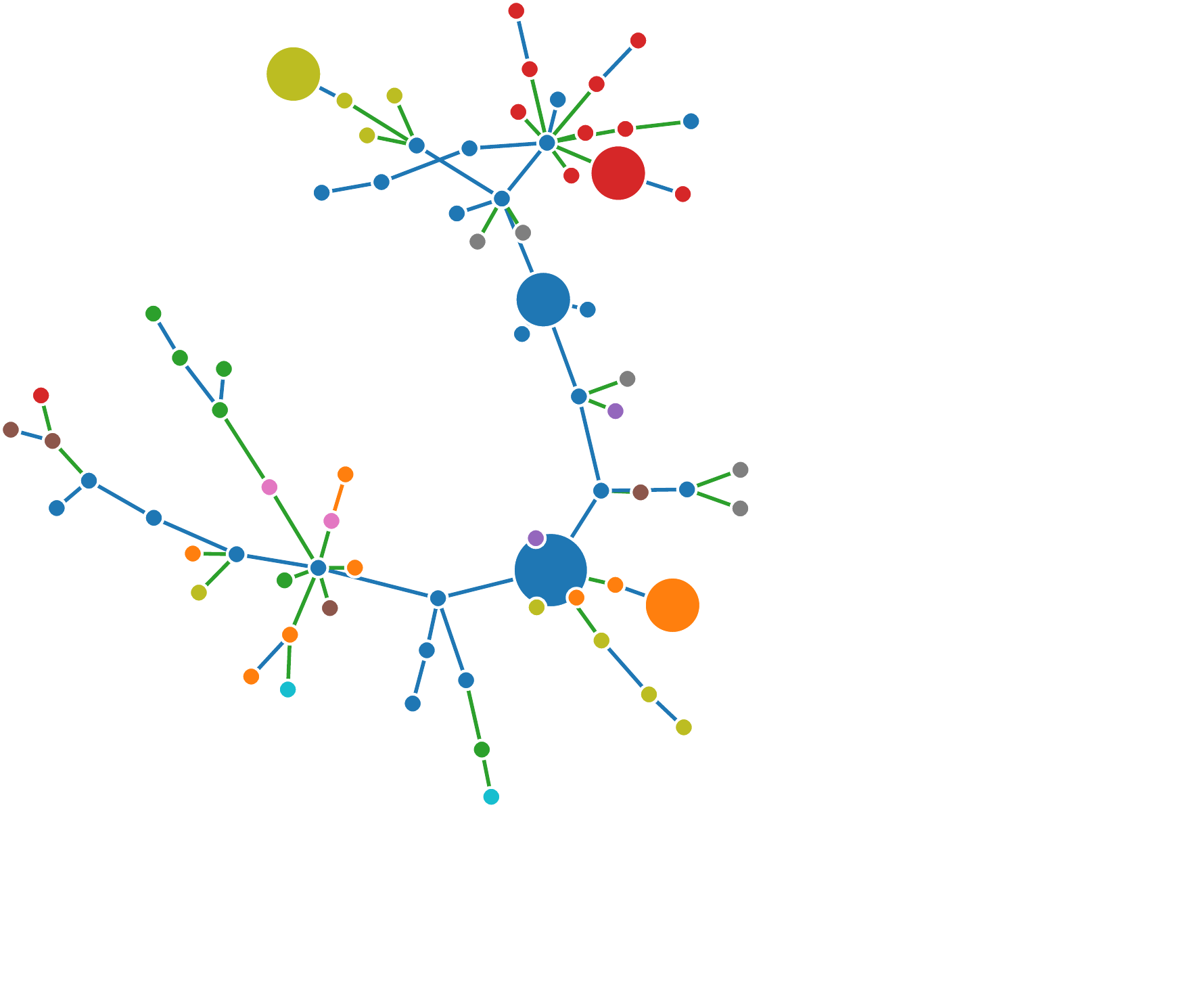}}    \put(0.55659602,0.08435243){\color[rgb]{0.12156863,0.46666667,0.70588235}\makebox(0,0)[b]{\textbf{\smash{government is against all kinds of roasts. }}}}    \put(0.55781109,0.04221208){\color[rgb]{0.83921569,0.15294118,0.15686275}\makebox(0,0)[b]{\textbf{\smash{why stop at beef? u should ban milk too.}}}}    \put(0.62243746,0.00038948){\color[rgb]{0.7372549,0.74117647,0.13333333}\makebox(0,0)[b]{\textbf{\smash{those for \#beefban should stop wearing leather}}}}    \put(0.50821266,0.12961428){\color[rgb]{1,0.49803922,0.05490196}\makebox(0,0)[b]{\textbf{\smash{this is why bjp will never win in kerala}}}}  \end{picture}\endgroup   \label{main:a}
}
\end{minipage}\begin{minipage}{.5\linewidth}
\centering
\subfloat[]{
    \def\svgwidth{\columnwidth}
    \begingroup  \makeatletter  \providecommand\color[2][]{    \errmessage{(Inkscape) Color is used for the text in Inkscape, but the package 'color.sty' is not loaded}    \renewcommand\color[2][]{}  }  \providecommand\transparent[1]{    \errmessage{(Inkscape) Transparency is used (non-zero) for the text in Inkscape, but the package 'transparent.sty' is not loaded}    \renewcommand\transparent[1]{}  }  \providecommand\rotatebox[2]{#2}  \ifx\svgwidth\undefined    \setlength{\unitlength}{479.24375bp}    \ifx\svgscale\undefined      \relax    \else      \setlength{\unitlength}{\unitlength * \real{\svgscale}}    \fi  \else    \setlength{\unitlength}{\svgwidth}  \fi  \global\let\svgwidth\undefined  \global\let\svgscale\undefined  \makeatother  \begin{picture}(1,0.88053399)    \put(0,0){\includegraphics[width=\unitlength]{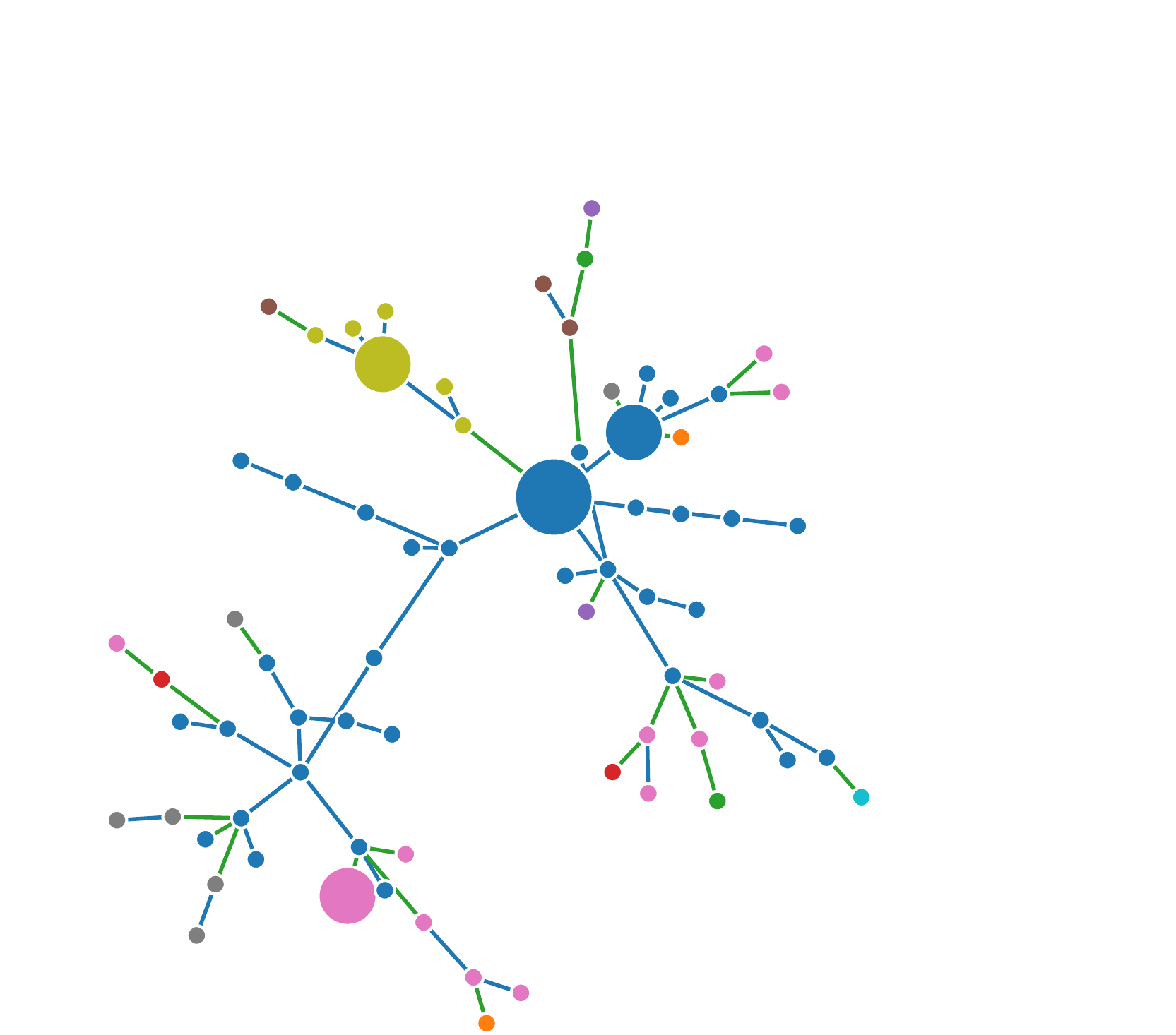}}    \put(0.25491556,0.80014692){\color[rgb]{0.89019608,0.46666667,0.76078431}\makebox(0,0)[b]{\textbf{\smash{why beef and not rice? both are living organisms.}}}}    \put(0.20979721,0.8602416){\color[rgb]{0.7372549,0.74117647,0.13333333}\makebox(0,0)[b]{\textbf{\smash{sir thanks 4 \#beefban! u hv proven u can take tough step}}}}    \put(0.30661641,0.75006802){\color[rgb]{0.12156863,0.46666667,0.70588235}\makebox(0,0)[b]{\textbf{\smash{good news: beef banned in maharashtra, 5 yrs jail}}}}  \end{picture}\endgroup   \label{main:b}
}
\end{minipage}
\caption{Extracted events for \beefban.  (a) 1st event demonstrates sign of opinion propagation.  
(b) 2nd event containing mixed opinions. Nodes are colored by the senders. The largest node is the root. }
\label{fig:beefban}
\end{figure}

\spara{\beefban.}
For the controversial ``beef ban'' \footnote{http://indianexpress.com/article/explained/explained-no-beef-nation/} law in India, 
results demonstrate clear separation of opinions among events. In Fig.~\ref{fig:beefban}, the 1st and 2nd event represents opinions opposing and supporting the law. However, we are not able to interpret any temporal pattern in the events due to the short timespan (3 days). 
We also observe the following. 
First, certain event (Fig. \ref{fig:beefban} (a)) display evidence of information propagation. For example, opposing opinions spreads along the user network and affected users also express their objection. 
Second, for some event (Fig.~\ref{fig:beefban}  (b)), dominant user exists who sent more than half of the tweets. 
Third, we observe events with mixed opinions (Fig. \ref{fig:beefban}  (b)). 
Last, our method tends to discover events at the ``peak'' as the set cover objective is better than the ``bottom'' (Fig.~\ref{fig:stream_graphs} (b)). 

\spara{\baltimore.}
We discovered two types of events: (1)  ``emotional'' events showing anger towards the riot, (2) ``descriptive'' events reporting current situation.

 \section{Conclusions}
\label{sec:conc}

We defined the problem of summarizing top-\nevents events in an interaction network. 
Our approach consists by first transforming the input data into an \metagraph 
and then defining two optimization problems: 
budgeted version of PCST and maximum set cover. 
We offer three algorithms for the former problem. 
Our experiments show that the greedy approach is more lightweight and performs as good as 
or even better than other more sophisticated counterparts. 

Our work opens many interesting directions for future research. 
For example, it would be interesting to formulate the problem differently, for example, imposing edge weight constraint for each edge, instead of their weight sum. 
Another direction is to explore semi-structured interactions such as forums posts, 
where nesting structures exist between post and comment. 
We leave scalability experiment and better summarization techniques for future work.

 
\spara{Acknowledgements.}
This work is partially supported by the Academy of Finland project ``Nestor'' (286211) 
and the EC H2020 RIA project ``SoBigData'' (654024). 

\bibliographystyle{plain-initials}
\bibliography{bibliography}

\begin{thebibliography}{10}

\bibitem{becker2011beyond}
H.~Becker, M.~Naaman, and L.~Gravano.
\newblock Beyond trending topics: Real-world event identification on twitter.
\newblock In {\em ICWSM}, 2011.

\bibitem{blei2003latent}
D.~Blei, A.~Ng, and M.~Jordan.
\newblock Latent dirichlet allocation.
\newblock {\em Journal of Machine Learning Research}, 3:993--1022, 2003.

\bibitem{boden2012mining}
B.~Boden, S.~G{\"u}nnemann, H.~Hoffmann, and T.~Seidl.
\newblock Mining coherent subgraphs in multi-layer graphs with edge labels.
\newblock In {\em KDD}, 2012.

\bibitem{breunig2000lof}
M.~Breunig, H.-P. Kriegel, R.~Ng, and J.~Sander.
\newblock {LOF:} identifying density-based local outliers.
\newblock {\em SIGMOD record}, 29(2), 2000.

\bibitem{cataldi2010emerging}
M.~Cataldi, L.~Di~Caro, and C.~Schifanella.
\newblock Emerging topic detection on twitter based on temporal and social
  terms evaluation.
\newblock In {\em IWMDM}, 2010.

\bibitem{celikyilmaz2011discovery}
A.~Celikyilmaz and D.~Hakkani-T{\"u}r.
\newblock Discovery of topically coherent sentences for extractive
  summarization.
\newblock In {\em ACL}, 2011.

\bibitem{charikar1999approximation}
M.~Charikar, C.~Chekuri, T.~Cheung, Z.~Dai, A.~Goel, S.~Guha, and M.~Li.
\newblock Approximation algorithms for directed steiner problems.
\newblock {\em J. of Algorithms}, 1999.

\bibitem{fiore2004uc}
A.~Fiore and J.~Heer.
\newblock {UC Berkeley Enron email analysis}, 2004.

\bibitem{garimella2015quantifying}
K.~Garimella, G.~De~Francisci~Morales, A.~Gionis, and M.~Mathioudakis.
\newblock Quantifying controversy in social media.
\newblock In {\em WSDM}, 2016.

\bibitem{guralnik1999event}
V.~Guralnik and J.~Srivastava.
\newblock Event detection from time series data.
\newblock In {\em KDD}, 1999.

\bibitem{inouye2011comparing}
D.~Inouye and J.~K. Kalita.
\newblock Comparing twitter summarization algorithms for multiple post
  summaries.
\newblock In {\em PASSAT and SocialCom}, 2011.

\bibitem{johnson2000prize}
D.~Johnson, M.~Minkoff, and S.~Phillips.
\newblock The prize collecting {S}teiner tree problem: theory and practice.
\newblock In {\em SODA}, 2000.

\bibitem{kumar2010dynamics}
R.~Kumar, M.~Mahdian, and M.~McGlohon.
\newblock Dynamics of conversations.
\newblock In {\em KDD}, 2010.

\bibitem{leskovec2009meme}
J.~Leskovec, L.~Backstrom, and J.~Kleinberg.
\newblock Meme-tracking and the dynamics of the news cycle.
\newblock In {\em KDD}, 2009.

\bibitem{mathioudakis2010Twittermonitor}
M.~Mathioudakis and N.~Koudas.
\newblock Twittermonitor: trend detection over the twitter stream.
\newblock In {\em SIGMOD}, 2010.

\bibitem{meladianos2015degeneracy}
P.~Meladianos, G.~Nikolentzos, F.~Rousseau, Y.~Stavrakas, and M.~Vazirgiannis.
\newblock Degeneracy-based real-time sub-event detection in twitter stream.
\newblock In {\em ICWSM}, 2015.

\bibitem{vazirani2013approximation}
V.~Vazirani.
\newblock {\em Approximation algorithms}.
\newblock Springer, 2013.

\bibitem{weng2011event}
J.~Weng.
\newblock Event detection in twitter.
\newblock In {\em ICWSM}, 2011.

\bibitem{yang2010modeling}
J.~Yang and J.~Leskovec.
\newblock Modeling information diffusion in implicit networks.
\newblock In {\em ICDM}, 2010.

\bibitem{yang2011patterns}
J.~Yang and J.~Leskovec.
\newblock Patterns of temporal variation in online media.
\newblock In {\em WSDM}, 2011.

\end{thebibliography}
\appendix
\section{NP-hardness of \pcst}
\label{sec:apx}

We prove \pcst is \np-hard. 
We induce \pcst from \textit{minimum set cover} problem. 

We define \textit{minimum set cover} as a decision problem, $SC(S, \mathcal{C}, k)$. 
Given $S=\set{s_1, \dots, s_n}, \mathcal{C}=\set{C_1, \ldots, C_m}$,  $C_i \subseteq S, i=1,\ldots,m$, 
$k \in \mathbb{Z}$, 
we are asked if there is $\mathcal{C}^{'} \subseteq \mathcal{C}$ 
such that $\forall s \in S, s \in \cup_{C \in \mathcal{C}^{'}} C$ and
$\abs{\mathcal{C}^{'}} \le k$.

\pcst is essentially a special case of budget version of PCST, where the input graph is DAG and node weight is 1. 
We denote the decision version of \pcst as $\pcsd(G, r, B, W)$. 
Given directed graph $G=(V, E)$, $c: E \rightarrow \mathbb{R}$, 
$\nodeweight: V \rightarrow 1$, 
$\treeroot \in V$, 
$\distbudget \in \mathbb{R}$ and $W \in \mathbb{R}$, 
we are asked if there exists a subtree $\event = (V^{'}, E^{'}) \subseteq  \mgraph$ rooted at $\treeroot$
such that $ c(\event) = \sum\limits_{e \in E^{'}} c(e) \le \distbudget$ and
 $\nodeweight(\event) = \sum\limits_{v \in V^{'}} \nodeweight(e) \ge W$.

Consider $SC(S, \mathcal{C}, k)$, 
where $S=\set{s_1, \ldots, s_n}, \mathcal{C}=\set{C_1, \ldots, C_m}$. 
Construct a directed graph $G=(V, E)$ where:

\begin{itemize} 
\item $V = \set{\treeroot, C_1, \ldots C_m, s_1, \ldots s_n}$
\item $E = \set{(\treeroot, C_i) \mid C_i \in \mathcal{C}} \cup \set{(C_i, s_j) \mid s_j \in C_i, C_i \in \mathcal{C}}$
\item $\forall C_i \in \mathcal{C}, \mweight(r, C_i)=N$
\item $\forall (C_i, s_j) \in \mathcal{C} \times S, \mweight(C_i, s_j)=1$
\item $N>n$
\end{itemize}

\begin{lemma}
	If $SC(S, \mathcal{C}, k)$ holds, $\pcsd(G, \treeroot, kN+n, k+n+1)$ holds. 
\end{lemma}

Without loss of generality, let $\mathcal{C}^{'} = \set{C_{q_1}, \ldots, C_{q_k}}$ be the set cover for $S$. 
We can construct a tree $T = (V^{'}, E^{'})$ such that 
$\nodeweight(T) > k+n+1$ 
and $c(T) \le kN+n$ in the following way:

\begin{itemize}
	\item $V^{'}= \set{\treeroot} \cup \mathcal{C}^{'} \cup S$
	\item $ (\treeroot, C_i) \in E^{'} $ if $C_i \in \mathcal{C}^{'}$
	\item $ (C_1, s_j) \in E^{'}$ if $s_j \in C_1$ 
	\item $ (C_i, s_j) \in E^{'}$ if $s_j \in C_i \setminus \bigcup\limits_{i=1 \ldots j-1} C_{q_i} $
\end{itemize}

It is easy to see $T$ makes $\pcsd(G, \treeroot, kN+n, k+n+1)$ hold. 

\begin{lemma}
	If $\pcsd(G, \treeroot, kN+n, k+n+1)$ holds, $ SC(S, \mathcal{C}, k) $ holds.
\end{lemma}

Denote the tree corresponding to $\pcsd(G, \treeroot, kN+n, k+n+1)$ as $T=(V^{'}, E^{'})$, 
then one set cover for $SC(S, \mathcal{C}, k)$ to hold is $\mathcal{C^{'}} = V^{'} \cap \mathcal{C}$. 
This can be proved by contradiction. 

On on hand, if $\mathcal{C^{'}}$ is not a set cover, then $\exists s \in S, s \not\in V^{'}$. 
Then some $C^{'} \in \mathcal{C}$ needs to be included in $\mathcal{C^{'}}$ to satisfy $p(T) \ge k+n+1$, . 
In this case $c(T) \ge (k+1)N > kN + n$, which is a contradiction. 

On the other hand, if $\abs{\mathcal{C^{'}}} > k$, then $c(T) \ge (k+1)N > kN + n$, which is a contradiction. 

Therefore, \pcst is \np-hard.

 
\end{document}